\documentclass[10pt,twocolumn,
 amsmath,amssymb,
 aps,
pra,
]{revtex4-2}

\usepackage{graphicx}
\usepackage{dcolumn}
\usepackage{bm}
\usepackage{xcolor}
\usepackage[flushleft]{threeparttable}
\usepackage{multirow}
\usepackage[colorlinks,linkcolor=blue,anchorcolor=blue,citecolor=blue,]{hyperref}
\newcolumntype{.}{D{.}{.}{-1}}
\newcolumntype{d}[1]{D{.}{.}{#1}}
\newcommand*{\wn}{cm$^{-1}$}
\newcommand*{\hsm}{H$_{2}$S}
\newcommand*{\Hm}{H$_{2}$}
\newcommand*{\X}{X$^1\Sigma_g^+$}

\newcommand*{\EF}{EF$^1\Sigma_g^+$}
\newcommand*{\E}{E$^1\Sigma_g^+$}
\newcommand*{\F}{F$^1\Sigma_g^+$}

\renewcommand{\eqref}[1]{Eq.~(\ref{#1})}

\begin{document}

\title{Precision measurement of the last bound states in H$_2$  \\ and determination of the H + H scattering length} 

\author{K.-F. Lai}
 \altaffiliation{Present Address: Department of Physics, The University of Hong Kong}
 \affiliation{Department of Physics and Astronomy, LaserLaB, Vrije Universiteit \\
 De Boelelaan 1081, 1081 HV Amsterdam, The Netherlands}
  \author{W. Ubachs}
  \email{w.m.g.ubachs@vu.nl}
  \affiliation{Department of Physics and Astronomy, LaserLaB, Vrije Universiteit \\
 De Boelelaan 1081, 1081 HV Amsterdam, The Netherlands}
 \author{M. Beyer}%
 \email{m.beyer@vu.nl}
 \affiliation{Department of Physics and Astronomy, LaserLaB, Vrije Universiteit \\
 De Boelelaan 1081, 1081 HV Amsterdam, The Netherlands}

\date{\today}

\begin{abstract}
The binding energies of the five bound rotational levels $J=0-4$ in the highest vibrational level $v=14$ in the \X\ ground electronic state of H$_2$ were measured in a three-step ultraviolet-laser experiment. 
Two-photon UV-photolysis of H$_2$S produced population in these high-lying bound states, that were subsequently interrogated at high precision via Doppler-free spectroscopy of the \F\ - \X\ system. 
A third UV-laser was used for detection through auto-ionizing resonances.
The experimentally determined binding energies were found to be in excellent agreement with calculations based on non-adiabatic perturbation theory, also including relativistic and quantum electrodynamical contributions.
The $s$-wave scattering length of the H + H system is derived  from the binding energy of the last bound $J=0$ level via a direct semi-empirical approach, yielding a value of $a_s$ = 0.2724 (5) $a_0$, in good agreement with a result from a previously followed theoretical approach. The subtle effect of the $m\alpha^4$ relativity contribution to $a_s$ was found to be significant.
In a similar manner a value for the $p$-wave scattering volume is determined via the $J=1$ binding energy yielding $a_p$ = -134.0000 (6) $a_0^3$.
The binding energy of the last bound state in H$_2$, the ($v=14$, $J=4$) level, is determined at 0.023 (4) \wn, in good agreement with calculation.
The effect of the hyperfine substructure caused by the two hydrogen atoms at large internuclear separation, giving rise to three distinct dissociation limits, is discussed.
\end{abstract}

\maketitle

\section{Introduction}

The $s$-wave scattering length for hydrogen represents a parameter governing many processes where collisions between hydrogen atoms at low temperatures play a role~\cite{Jamieson1992}.
Apart from its theoretical relevance, hydrogen being the simplest atom, the scattering length is of importance in many applications, such as in the recombination kinetics of hydrogen~\cite{Roberts1969}, also playing a role in the formation of primordial stars~\cite{Palla1983,Forrey2016}.
Further, it describes shifts in the resonance frequency in the H-maser~\cite{Wittke1956} as well as in precision measurements for the determination of the Rydberg constant in the H-atom~\cite{Jentschura2019,Matveev2019}. 
Both the ground state singlet $a_{1S-1S}$ (further denoted as $a_s$) as well as the $a_{1S-2S}$ scattering lengths play a role in the formation of a hydrogen Bose-Einstein condensate~\cite{Fried1998,Killian1998}.
The value of $a_s$ can be computed based on ab initio calculations of the diatomic potential energy curve~\cite{Gribakin1993,Szmytkowski1995,Sen2006a}.
However, subtle effects in such computations may have a large influence on the outcome. 
Nonadiabatic couplings are known to play a major role~\cite{Dalgarno1956,Wolniewicz2003b}, and a small change in mass, e.g. going from nuclear to atomic masses, may result in a large change in the value of $a_s$~\cite{Jamieson2010}. 

In a previous study by our group~\cite{Lai2021b,Lai2022} the H$_2$ potential curve as calculated by the advanced method of non-adiabatic perturbation theory (NAPT), including relativistic and high-order quantum electrodynamical (QED) contributions up to $m\alpha^6$~\cite{Czachorowski2018}, was tested in the energy range around the dissociation limit and at large internuclear separations. 
Good agreement was found between experimental and theoretical values for the level energies for high vibrational and rotational quantum numbers, demonstrating that the NAPT approach yields high ($v,J$) levels in H$_2$ at $10^{-3}$ \wn\ precision.
The experimentally verified NAPT-potential energy curve was used to derive values for the ground state scattering length, resulting in $a_s = 0.274\,(4)\,a_0$.
Importantly, the effects of introducing adiabatic, non-adiabatic, relativistic and QED corrections beyond the Born-Oppenheimer (BO) potential on the value of $a_s$ could be computed and large variations were indeed found upon including those subtle corrections. 
The final value obtained is a factor of two smaller than a value based only on the BO-potential ($a_s = 0.570$ $a_0$), while the inclusion of QED corrections were found to have an effect as large as 6\%~\cite{Lai2022}.

As an alternative different approaches exist for determining the scattering length directly from experimental data. 
The observation of minima in photoassociation spectra can be exploited to determine nodes in the collision wave function of cold atoms~\cite{Gardner1995}.
Spectroscopic data of bound vibrational levels can be used to determine phases of the last levels, which then yield accurate information on the scattering lengths, as was shown for Rb$_2$~\cite{Tsai1997} and Na$_2$~\cite{Crubellier1999}.
Also the observation of Feshbach resonances can provide scattering lengths as was shown for Cs$_2$~\cite{Lange2008f}. 
Recently, a semi-empirical scaling method, relating $a_s$ to experimental values of the binding energy of weakly bound dimers was applied~\cite{Augusticova2021}.
In this 'direct method', potential energy curves are parametrized.

For such a direct determination of the $s$-wave scattering length the level energy of the last bound $J=0$ is required as experimental input.
In the experimental part of the present study, an accurate measurement of the level energies is performed for all five ($J=0-4$) bound rotational levels in \X, $v=14$.
The experimental methods follow those of previous three laser-excitation studies, whereby H$_2$ molecules are produced in the highest vibrational level through two-photon ultraviolet (UV) photolysis, followed by two-photon precision measurements in the \F\ - \X\ system and a third laser excitation for ionization and detection~\cite{Trivikram2016,Trivikram2019,Lai2021,Lai2021b,Lai2022}.

\section{Experimental Methods}

The experimental setup and methods for the production and detection of highly excited rovibrational states in hydrogen has been presented in previous work ~\cite{Trivikram2016,Trivikram2019,Lai2021,Lai2021b,Lai2022,Lai2023}. The highest rovibrational states H$_2^*$($v$, $J$) for $v=14$ and $J=0-4$ are produced through two-photon UV-photolysis of~\hsm\ via the pathway~\cite{Steadman1989}:
\begin{equation*}
    \text{H$_{2}$S} \xrightarrow{2\lambda_{\rm UV}} \text{S} (^{1}{\rm D}_{2}) + \text{H$_{2}^{*}$}
\end{equation*}
The (pre)-dissociation of the parent \hsm\ molecules proceeds through the $3d\,^1$A$_1$ electronic state with the pulsed UV-laser fixed at $\lambda_{\rm UV}$ = 281.8 nm throughout all measurements. 
As discussed previously~\cite{Lai2023} this excitation has sufficient energy for full dissociation into S($^1$D$_2$) + 2 H(1s) leaving an excess energy of over 1000 \wn.
It is remarkable that in the photodissociation process producing S($^1$D$_2$) only 1.5\% of the energy is released into kinetic energy, while virtually all energy is spent on electronic and vibrational excitation of the products.
That is under the condition that singlet-triplet selection rules are obeyed, and that no symmetry breaking, possibly via spin-orbit interaction, or intra-molecular singlet-triplet relaxation in H$_2$S occurs before dissociation. 

\begin{figure}[!t]
\begin{center}
\includegraphics[width=\linewidth]{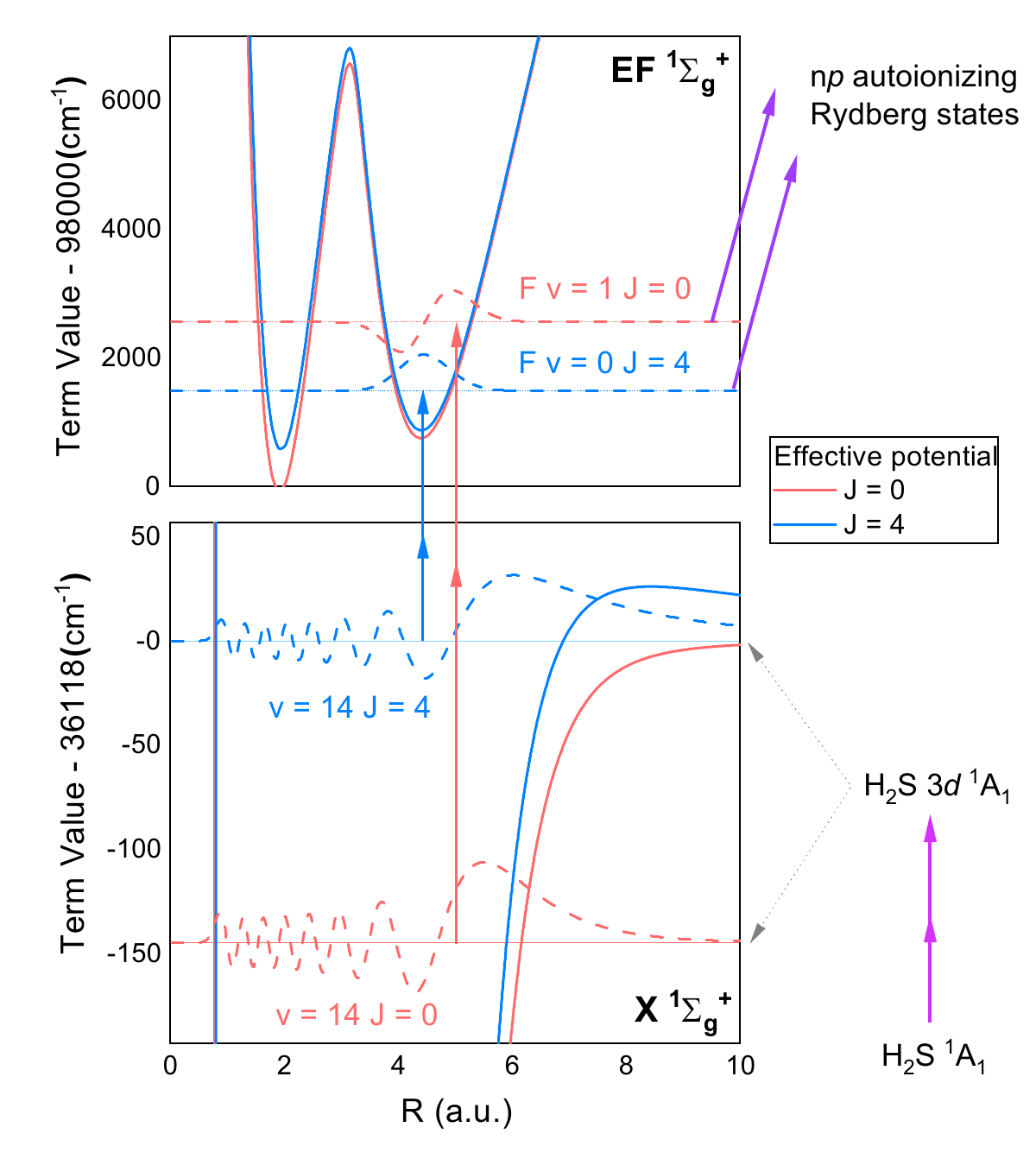}
\caption{\label{level_scheme}
Effective potential energy curves of \X\ and \EF\ states for $J=0$ (red) and $J=4$ (blue) including rotational energies. The wave functions for the highest vibrational states $v=14$ are plotted as dashed lines. Note that the $v=14, J=4$ wave function is well-extended beyond 10 a.u. The initial two-photon photolysis step producing H$_2^*$ and the final autoionization step are depicted at the right hand side in the figure.
}
\end{center}
\end{figure}

The H$_2^*$ states produced from UV photolysis are interrogated via two-photon Doppler-free laser spectroscopy, with counter-propagating laser beams in a Sagnac configuration~\cite{Hannemann2007}, for excitation into the \F\ $v=0$ (F0) or $v=1$ (F1) states. 
Further excitation of the F0 or F1 population into autoionizing Rydberg states with another UV laser, then produces H$_2^+$ ions for signal detection. For each \F\ - \X\ transition investigated the third laser is tuned to a strong autoionizing resonance.
The H$_2^+$ ions generated in the latter process were mass selected in a time-of-flight mass spectrometer and detected by a microchannel plate.
A level scheme displaying the three-step laser excitation process and some relevant $J$-dependent potential energy curves of the \X\ and \F\ states are presented in Fig.~\ref{level_scheme}.

The photolysis process produces only little population in the highest $v=14$ vibrational level, and in a previous study only the $v=14, J=1$ level could be detected via the F1-X14 Q(1) transition. 
The detection sensitivity of the experimental setup is now improved by applying a DC-field at about 1.3-1.4 kV/cm with reversed polarity before the arrival of the ionization laser pulse. 
The H$_2^+$ ions produced from the powerful photolysis laser  
were thereby removed, giving rise to a largely reduced background signal. 
The dc-Stark shift induced by this method amounts to less than 1 MHz and is negligible at the present measurement accuracy.

\section{Spectroscopic Results}

\begin{figure}[!t]
\begin{center}
\includegraphics[width=\linewidth]{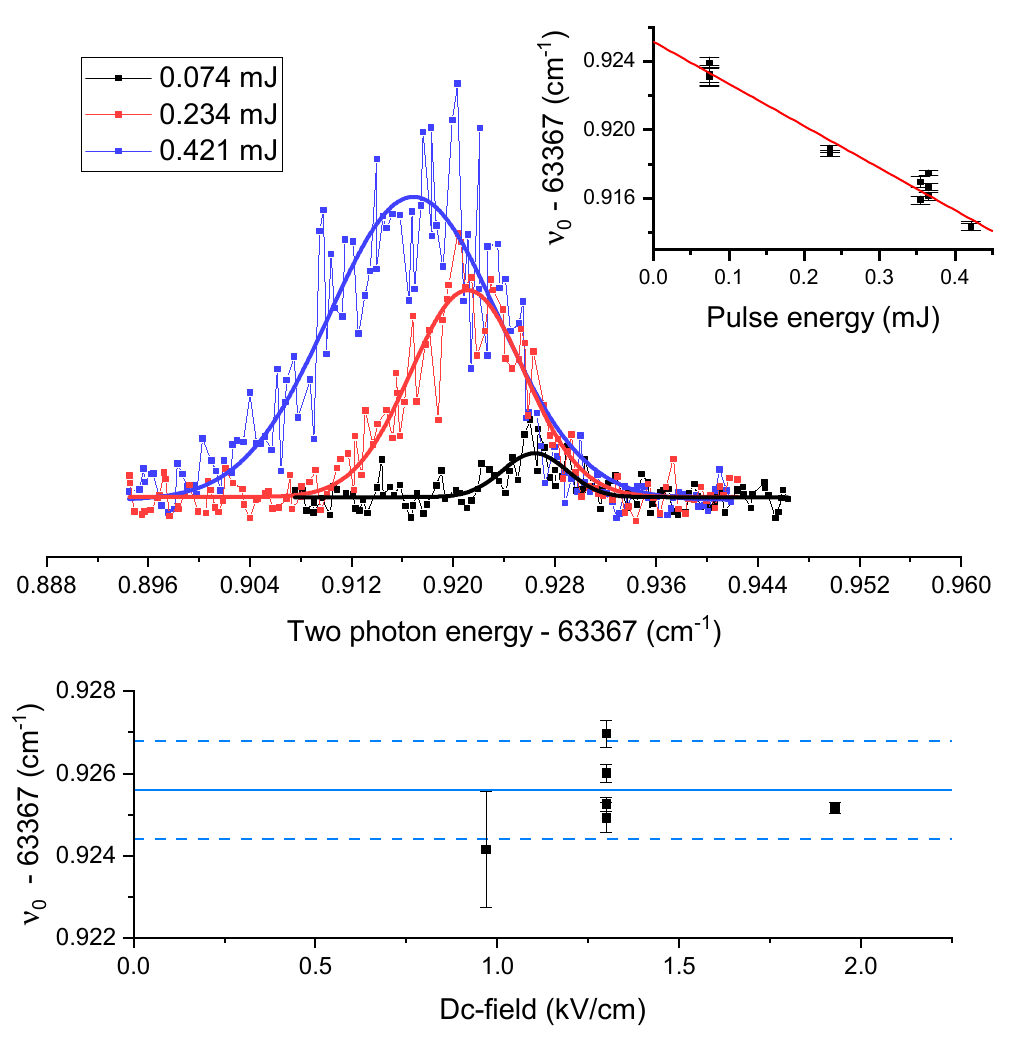}
\caption{\label{F0X14Q4}
Spectra of the F0-X14 Q(4) transition measured at different pulse energies in the presence of a 1.3 kV/cm reverse bias dc-field. Full lines represent fitted Voigt curves through the data. The upper inset shows the extrapolation to zero pulse energy/intensity. In the lower panel the transition frequencies obtained for different reverse bias dc-fields are plotted.
}
\end{center}
\end{figure}

\begin{table}
\renewcommand{\arraystretch}{1.3}
\caption{\label{tab:transition}
Measured frequencies for the F-X transitions with uncertainties indicated in parentheses. 
}
\begin{threeparttable}
\begin{tabular}{ccd{9}d{8}.}
	&  & \multicolumn{1}{c}{Frequency (\wn)}  \\
\hline
       & Q(0) & 64\,585.588 (5)    \\
F1-X14 & Q(1) & 64\,580.408 (4)^{a,b}  \\
       & Q(2) & 64\,571.676 (5)  \\
\hline
\multirow{2}{*}{F0-X14}  & Q(3) & 63\,369.334 (5)   \\
       & Q(4) & 63\,367.9248 (27)   \\
\hline
\hline
       & Q(0) & 65\,207.533 (5)   \\
F1-X13 & Q(1) & 65\,188.589 (5)  \\
       & Q(2) & 65\,151.813 (5)  \\
\hline
F0-X13 & Q(3) & 63\,905.9305 (24)^c \\
F0-X12 & Q(4) & 64\,781.6783 (11)^a   \\
F0-X11 & Q(4) & 66\,105.8614 (14)^a   \\
\hline
\end{tabular}
\begin{tablenotes}
\footnotesize
\item $^a$ Measured under dc-field free conditions.
\item $^b$ Revised value. See text for details.
\item $^c$ Ref. \cite{Lai2021}
\end{tablenotes}
\end{threeparttable}
\end{table}

Based on the sensitivity improvements, all five $J=0-4$ bound states in \X, $v=14$ of \Hm\ could be detected via Doppler-free two-photon transitions in the \F\ - \X\ electronic system. 
Sample spectra of the F0-X14 Q(4) transition, obtained for various pulse energies of the spectroscopy laser, are shown in Fig.~\ref{F0X14Q4}.
Center frequencies of the measured resonances are determined by fitting Voigt curves to the experimental data sets.
An ac-Stark shift-analysis with extrapolation to zero-pulse energy for this Q(4) line is also displayed. 
Such an analysis and extrapolation is carried out for all lines reported in the present study.

\begin{figure}[!b]
\begin{center}
\includegraphics[width=\linewidth]{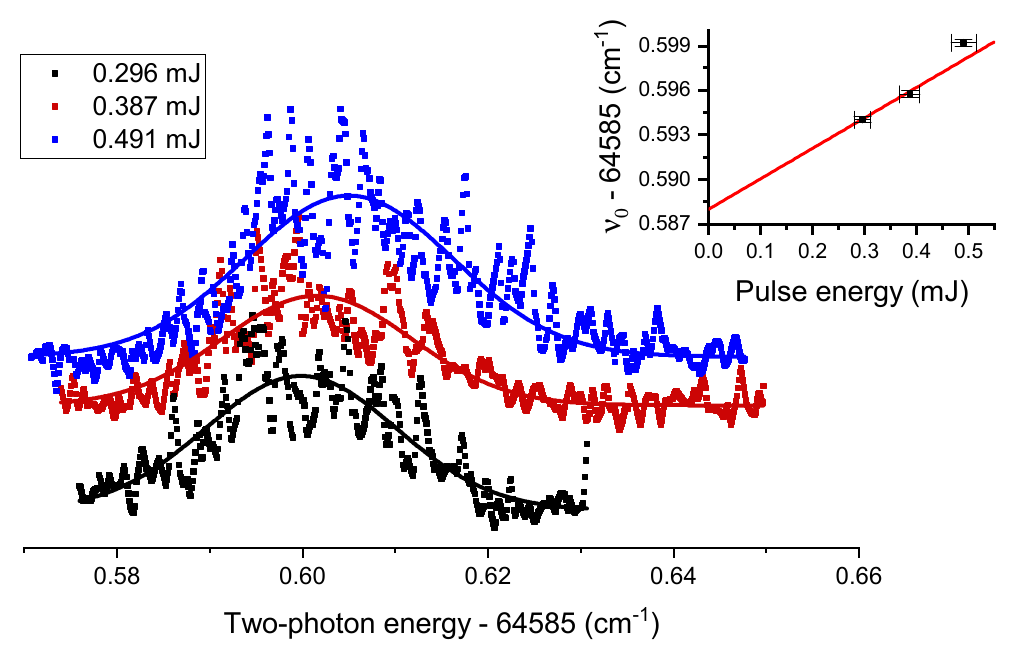}
\caption{\label{F1X14Q0}
Spectra of the F1-X14 Q(0) transition measured at different pulse energies in the presence of a 1.4 kV/cm reverse bias dc-field. Multiple scans ($\geq3$) are averaged to produce the spectra shown at the same pulse energies. The signal strength of spectra at different pulse energy should not be compared directly as the measurements were taken with different detection settings. Spectra are vertically translated for clarity.
}
\end{center}
\end{figure}

In Fig.~\ref{F1X14Q0} spectral recordings of a Q(0) line probing $v=14$, $J=0$ are displayed for various laser pulse energies. A special focus was given to measuring this level since the subsequent direct derivation of the scattering length depends on this specific result.
The signal in these spectra is very low even though the two-photon transition probes the F1-X14 band which exhibits a larger Franck-Condon factor than the F0-X14 band. The low signal strength is most likely due to the low population of the X(14,0) level in the photolysis process.
From a careful analysis and ac-Stark extrapolation the transition frequency could be determined at an uncertainty of 0.005 \wn.

The measured transition frequencies and their uncertainties are listed in Table \ref{tab:transition}. 
The \Hm\ $v=14, J=0-2$ levels are detected through F1-X14 Q-branch transitions.
The measurements of the Q(1) line in this band, the strongest line observed, had been presented in a previous publication as the only line probing a $v=14$ level~\cite{Lai2021}. 
This F1-X14 Q(1) transition is revisited and it is established that the previously reported value has an offset by -0.01~\wn, caused by the mis-assignment of the I$_2$ saturated absorption line used for the absolute frequency calibration of F1-X14 Q(1) spectra~\cite{Iospec}. 
The signal strength of the F1-X14 Q(2) transition is similarly weak as that of Q(0), shown in Fig.~\ref{F1X14Q0}.

The $J=3$ and 4 levels in the $v=14$ ground state were probed via Q(3) and Q(4) transitions in the  F0-X14 band, where Q(4) has the strongest signal strength.
A dc-Stark analysis is performed on the F0-X14 Q(4) line at fixed laser pulse energy, results of which are shown in Fig.~\ref{F0X14Q4}.
From these data we conclude that a dc-field of up to 1.3-1.4 kV/cm has no observable effect on the spectroscopic measurements under the current conditions.
As discussed previously, this finding is in agreement with calculations based on polarisabilities
for the EF state~\cite{Komasa2005} and X state~\cite{Kolos1967}, predicting a less than 1 MHz shift for the dc-field used.
Most of the measurements were therefore carried out with a reverse bias field of 1.3 kV/cm, therewith enhancing the signal strength and signal-to-noise ratio.

\begin{figure}[!b]
\begin{center}
\includegraphics[width=\linewidth]{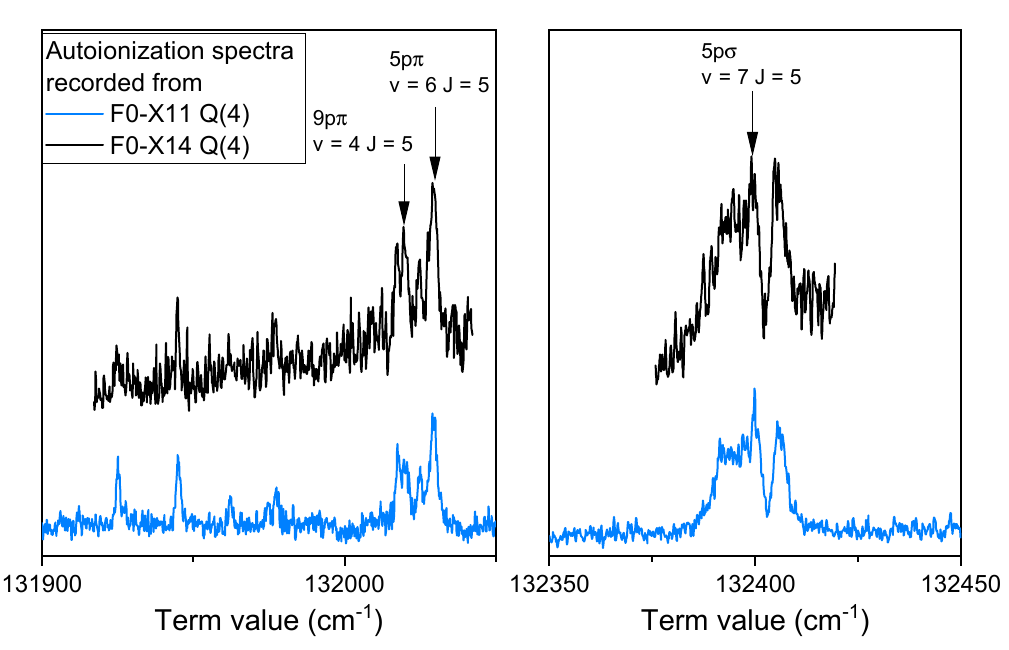}
\caption{\label{autoionization}
Portions of autoionization spectra recorded from the F0 $J=4$ intermediate level populated through F0-X14 Q(4) (in black) and F0-X11 Q(4) (in blue) two-photon transitions.
}
\end{center}
\end{figure}

In order to compare with theoretical calculations, further measurements are performed probing $v=11-13$ levels in the \X\ ground electronic state of~\Hm, under the same experimental conditions.
Transition frequencies of F1-X13 Q(0-2) lines are listed in Table~\ref{tab:transition}. 
The field-free transition frequency of F0-X13 Q(3)~\cite{Lai2021} has also been included. 
The revisited F0-X11 Q(4) line and a newly measured F0-X12 Q(4) were recorded under dc-field free conditions. 

The signal strengths of the F-X two-photon transitions are boosted by setting the ionization laser at a wavelength matching an autoionization resonance. 
The assignments of the two-photon transitions in the F-X system could be verified by scanning the wavelength of the third laser inducing autoionization. 
Figure~\ref{autoionization} presents autoionization spectra originating in the F0 $J=4$ level, which was populated through F0-X11 Q(4) and F0-X14 Q(4) transitions. 
In both cases the autoionization was probed in the energy ranges (representing the total energy above the \X, $v=0$, $J=0$ ground level)
131\,900-132\,050 \wn\ and 132\,350-132\,450 \wn.
Most key autoionization features are duplicated via both excitation channels, similarly as was done in a previous study~\cite{Trivikram2019}, thus confirming the assignment of the X and F levels involved in the stepwise excitation pathways.

The error budget of the precision measurements in the three-step laser investigation was already discussed in a previous study probing X, $v=13,14$ levels~\cite{Lai2022}.
Minor contributions stem from the lineshape fitting, the absolute frequency calibration against the I$_2$ Doppler-free standard, and  the determination of cw-pulse frequency offset, or the frequency chirp of the pulsed dye amplifier system, amounting to a subtotal uncertainty of 25 MHz. 
A major uncertainty is associated with the ac-Stark extrapolation, which varies with individual transitions. 
For the weak transitions, F1-X14 Q(0,2), F0-X14 Q(3) and F1-X13 Q(0,2), measurements could only be performed at high pulse energies ($>0.1$ mJ), leading to a 150 MHz uncertainty in total.
The stronger F0-X14 Q(4), F0-X11 Q(4) and F0-X12 Q(4) transitions allow for measurements at lower laser pulse energies and better statistical averaging. 
The total uncertainties of the stronger transitions are as low as 30-90 MHz.
In Table~\ref{tab:transition} these values are converted into units of \wn.

\begin{figure}
\begin{center}
\includegraphics[width=\linewidth]{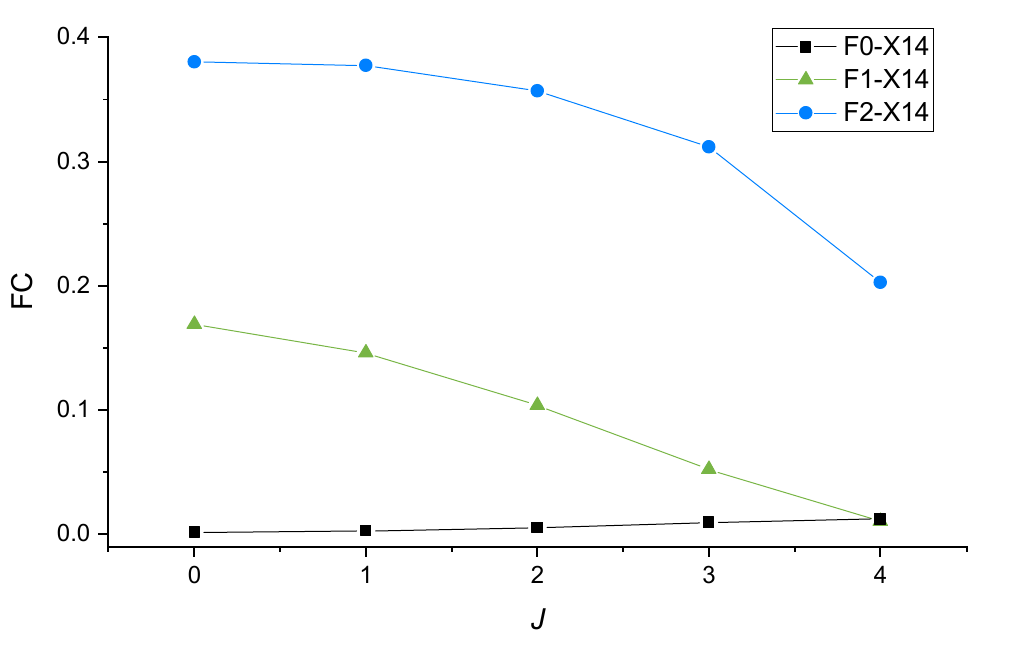}
\caption{\label{FC_F-X14}
Calculated Franck-Condon factors of F$v'$ - X14 Q-branch transitions.
}
\end{center}
\end{figure}

The relative intensities observed in the three-step excitation sequence involving the two-photon precision measurements in the \F\ - \X\ system are seemingly erratic.
Calculations of Franck-Condon factors (FCF), as shown in Fig.~\ref{FC_F-X14}, reveal that the FCF in the F2-X14 band are larger than those in F1-X14, while the FCF in F0-X14 are even smaller.
However, the F2-X14 progression is only observed in high-power low-resolution measurements~\cite{Lai2023}, while the F1-X14 Q(3) transition is absent despite of the 5-times larger FC for F1-X14 Q(3) than F0-X14 Q(3).
The FCF-values form insufficient ground for explaining the signal strengths in the three-color experiments. 
The population distribution produced in the photolysis process~\cite{Zhao2021,Zhao2022} as well as the  transition rates to autoionizing Rydberg states at large internuclear separations~\cite{Trivikram2019} are decisive factors for the amount of H$_2^+$ signal produced.

\section{Verification of combination differences in \X}

\begin{table}
\renewcommand{\arraystretch}{1.3}
\caption{\label{tab:com_diff}
Energy intervals determined by combination difference of measured transition presented in this work and comparison with theoretical values obtained via the NAPT program~\cite{SPECTRE2022}. All values are presented in \wn, with uncertainties indicated in parentheses.}
\begin{tabular}{cccc}
  Interval   &  This work  & Calculation~\cite{SPECTRE2022}  & Difference     \\
\hline
X14-X13 Q(0) & 621.945 (7) & 621.9582 (18) & -0.013 (7)\\
X14-X13 Q(1) & 608.176 (6) & 608.1816 (18) & -0.006 (6)\\
X14-X13 Q(2) & 580.137 (7) & 580.1353 (19) & 0.001 (7)\\
X14-X13 Q(3) & 536.597 (7) & 536.6001 (20) & -0.003 (7)\\
\hline
X14-X11 Q(4) & 2737.9358 (16) & 2737.9402 (42) & -0.0044 (46)\\
X14-X12 Q(4) & 1413.7527 (18) & 1413.7575 (35) & -0.0048 (39)\\
X12-X11 Q(4) & 1324.1831 (18) & 1324.1827 (8) & -0.0003 (20)\\
\hline
\end{tabular}
\end{table}

The consistency of the precision spectroscopic measurements of the F-X transitions can be verified by the determination of experimental combination differences in the \X\ ground state, which then can be compared with results of ab initio calculated values obtained via the most accurate code, the H2SPECTRE program~\cite{SPECTRE2022,Czachorowski2018} based on non-adiabatic perturbation theory.
For $J=0-3$, the vibrational intervals between $v=13$ and $v=14$ (X14-X13 Q($J$)) are determined and presented in Table~\ref{tab:com_diff}. 
For $J=4$, three intervals X14-X11, X14-X12 and X12-X11 are determined. 
The revisited F0-X11 Q(4) transition shows a -0.0081 \wn\ discrepancy with previous measurement, which is ascribed to a calibration error in Ref.~\cite{Trivikram2016}.  
Table~\ref{tab:com_diff} compares the experimental values with the calculated values for the ground state vibrational intervals, including uncertainties, as obtained from the H2SPECTRE  program. 
The experimental results agree well with their theoretical counterpart values except for the $J=0$ level, which presents a difference close to 2$\sigma$.
These findings confirm the conclusion of the previous study~\cite{Lai2021} that the NAPT formalism describes the highly excited H$_2$ rovibrational levels well, at the accuracy of experimental precision.

\section{Extraction of binding energies}

In the previous section a consistent picture is built that 
provides a verification for the binding energies of the X, $v=14$ levels through comparison with the theoretical NAPT formalism. Table~\ref{tab:com_diff} tests the computations via energy differences.
However, it is the goal of the present study to extract the binding energies via a purely experimental approach.
In high-resolution Fourier-transform emission spectroscopic studies, combined with Doppler-free two-photon laser studies, the rovibrational manifolds of nine different electronic states, including the \E\ inner well as the \F\ outer well states, were connected to the \X\ manifold and absolute term values were determined~\cite{Salumbides2008,Bailly2010}.

\begin{table*}
\caption{\label{tab:binding_erg}
Binding energies of all five bound states for $J=0-4$ in $v=14$ extracted directly from experiment (see text for details of analysis). The conversion of excitation energies of X14, $J$ into binding energies relies on the value of the dissociation energy $D_0 = 36\,118.069\,632\,(26)$ \wn~\cite{Beyer2019}. All values are presented in \wn, with uncertainties indicated in parentheses. 
}
\begin{tabular}{lllccc}
  Transition   &  Frequency    &F0/F1 energy~\cite{Bailly2010}  & X14 Excitation & X14 Binding & X14, $J$\\
  \hline
F1-X14\,Q(0)  & 64585.588 (5) & 100558.851 (1) & 35973.263 (5) & 144.807 (5) & 0 \\
F1-X14\,Q(1)  & 64580.408 (4) & 100570.8430 (3) & 35990.435 (4) & 127.635 (4)   & 1 \\
F1-X14\,Q(2)  & 64571.676 (5) & 100594.8070 (6) & 36023.131 (5) & 94.938 (5) &  2 \\
F0-X14\,Q(3)  & 63369.334 (5) & 99437.1665 (5) & 36067.832 (5) & 50.238 (5)  & 3  \\
F0-X14\,Q(4)  & 63367.9248 (27)  &  99485.972 (3) & 36118.047 (4) & 0.023 (4) & 4  \\
\hline
\end{tabular}
\end{table*}

Combining the presently measured transition frequencies connecting the  \X,$v=14$ levels to F0 and F1 outer well rovibrational levels (see Table~\ref{tab:transition}), with the term values of F0($J$) and F1($J$) from Bailly et al.~\cite{Bailly2010}, then results in values for the excitation energies of \X,$v=14$ levels above the \X,$v=0$, $J=0$ ground level of the molecule.
Further combining these results with the accurate value for the dissociation energy of H$_2$~\cite{Beyer2019}, $D_0 = 36\,118.069\,632\,(26)$ \wn, yields values for the binding energies of \X,$v=14$ levels, as listed in Table~\ref{tab:binding_erg}.
The overall uncertainty of experimentally determined binding energies amounts to 0.005 \wn, limited by the accuracy of the present measurements. 

\begin{figure}
\begin{center}
\includegraphics[width=\linewidth]{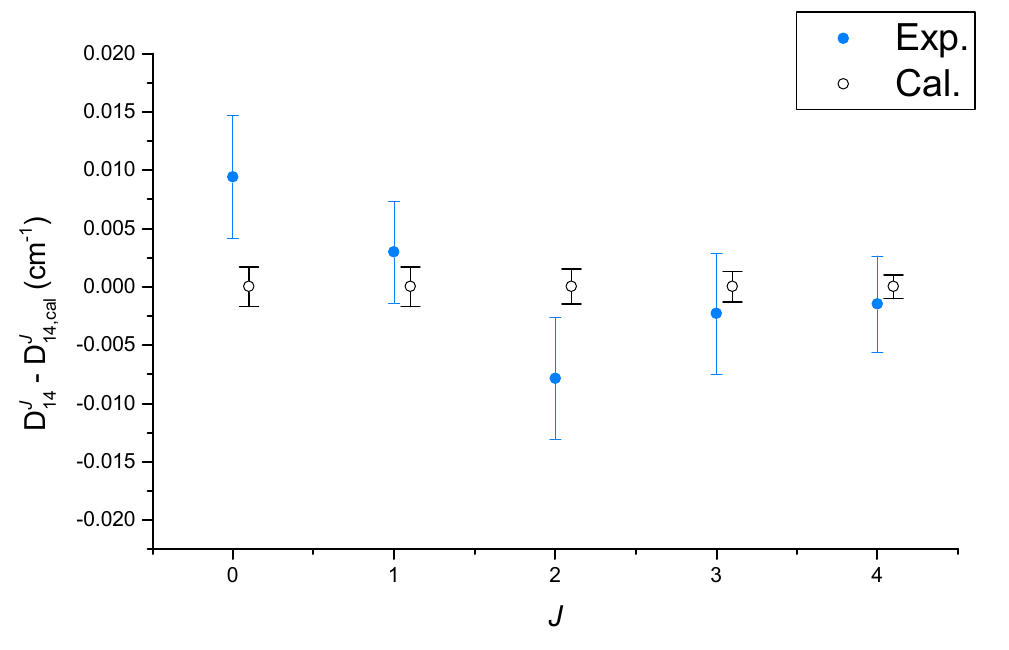}
\caption{\label{binding_erg}
The experimentally determined binding energies of all five bound states in \X, $v=14$, and a comparison with values derived from the NAPT ab initio approach~\cite{Czachorowski2018,SPECTRE2022}.
}
\end{center}
\end{figure}

The same values can be computed via the NAPT approach, coded in the H2SPECTRE program~\cite{SPECTRE2022,Czachorowski2018}.
The comparison between experimental and theoretical values is graphically shown in Fig.~\ref{binding_erg}.
The experimental values show good agreement with those from ab initio calculations, with combined uncertainties all within 1.5$\sigma$.

\section{$s$-wave scattering Length}
\label{swave}

In previous studies on high-lying bound and quasibound resonances, the experimentally observed rovibrational intervals verified the accuracy of the potential energy curve of \Hm\ in the NAPT formalism at large internuclear distances. Based on this verified potential the $s$-wave scattering length was computed yielding $a_s= 0.274 (4)$  $a_0$~\Hm~\cite{Lai2022}.

With the current measurement of the binding energy of the last bound $J=0$ level 
($D_{v_{\text{max}}}$) the $s$-wave scattering length can be computed in a direct manner through~\cite{landau2008quantum}:
\begin{equation}
    a_s = \sqrt{\frac{\hbar^2}{2\mu D_{v_{\text{max}}}}}.
    \label{as-eq}
\end{equation}
This procedure yields a value of $a_s=  0.908$  $a_0$ using the atomic reduced mass to include non-adiabatic effects in an approximate manner~\cite{Jamieson2010}.
It must be noted however that Eq.~(\ref{as-eq}) holds in the approximation of a purely $R^{-6}$ Van der Waals potential at large $R$, not being exact for H$_2$.

In the following we adopt a direct procedure for determining $a_s$, that uses the experimental binding energy of the last bound $J=0$ state in combination with a more realistic approach for the H$_2$ potential.
The nonadiabatic perturbation theory (NAPT) approach~\cite{Czachorowski2018} is employed to represent the H$_2$ potential curve.
Such perturbative approach allows to test the correlation of $D^J_{14}$  versus $a_J$ at different level of approximation including the Born-Oppenheimer (BO), adiabatic (AD), nonadiabatic (NA), relativistic terms and the QED contributions up to the $m\alpha^6$ term. 

The detailed NAPT calculation scheme of bound state level energies has been discussed in Ref. \cite{Komasa2019}. The nonrelativistic energy contribution within NAPT is evaluated by solving the radial nuclear Schr\"odinger equation:
\begin{align}
    \left[-\frac{1}{R^{2}} \frac{\partial}{\partial R} \frac{R^{2}}{2 \mu_{\|}(R)} \frac{\partial}{\partial R}+\frac{J(J+1)}{2 \mu_{\perp}(R) R^{2}}+f\mathcal{V}(R)\right] \phi_{i}(R) \notag\\
=E_i \phi_{i}(R),
\end{align}
with
\begin{align}
    \mathcal{V}(R) & = \mathcal{E}_\text{BO}(R) + \mathcal{E}_\text{AD}(R) + \delta\mathcal{E}_\text{NA}(R),
\end{align}
representing the BO potential energy curve, adiabatic correction and leading order nonadiabatic correction.
In this approach a scaling factor $f$ is introduced, following Ref.~\cite{Augusticova2021} that can later be determined in a fitting procedure to extract $a_s$.

The vibrational reduced mass and the rotational reduced mass are defined: 
\begin{align}
    \frac{1}{2\mu_{\|}(R)} & = \frac{1}{2\mu_\text{a}}+ W_{\|}(R),  \\
    \frac{1}{2\mu_{\perp}(R)} & = \frac{1}{2\mu_\text{a}}+ W_{\perp}(R),
\end{align}
where $\mu_a$ is the reduced atomic mass. 
The radial nuclear Schr\"odinger equation can be rewritten, without a first derivative part, by using the ansatz $\chi_i(R)=R\phi_i(R)\exp{[-Z(R)]}$:
\begin{align}\label{eq:SE_transf}
\left[-\frac{1}{2\mu_{\|}(R)} \frac{d^2}{d\,R^2} - \frac{\mu_{\|}(R)}{2} \left(W_{\|}'(R) \right)^2 + \frac{1}{2}W_{\|}''(R) \right.\notag\\
\left.+ \frac{1}{R}W_{\|}'(R) + \frac{J(J+1)}{2 \mu_{\perp}(R) R^{2}}+f\mathcal{V}(R)\right] \chi_{i}(R) =E_i \chi_{i}(R).
\end{align}
All the $R$-dependent potential terms and $W^{(i)}_{\|}(R)$ and $ W_{\perp}(R)$ functions are extracted from Ref.~\cite{Komasa2019}. 
The radial nuclear Schr\"odinger equation can be solved using the renormalized Numerov method~\cite{Johnson1977}.

The relativistic and QED corrections can be evaluated from the BO nuclear wave function obtained from the radial nuclear Schr\"odinger equation:
\begin{align}
\left[-\frac{1}{2\mu_{n}} \frac{d^2}{d\,R^2}
+ \frac{J(J+1)}{2 \mu_{n} R^{2}}+f\mathcal{E}_\text{BO}(R)\right] \chi_{i}(R) =E_i \chi_{i}(R).
\end{align}

The binding energies for different rovibrational levels were evaluated with the potential energy functions in the range from 0.001 to 50 a.u. with stepsize of 0.001 a.u. It showed good agreement with the values calculated using the H2SPECTRE program for $f=1$ unscaled potentials.

The scattering parameter can be evaluated from the radial wavefunction at zero energy ($k=\sqrt{2\mu_a(E-\mathcal{V}(\infty)}\to 0$). 
The asymptotic form $R\to\infty$ of the radial wavefunction at given $J$ is:
\begin{equation}
    \lim_{\substack{R\to\infty \\ k\to 0}}\chi(R;k) \propto kR(j_{J}(kR)\cos\eta_{J}(k) - n_{J}(kR)\sin\eta_{J}(k)) ,  
\end{equation}
where $j_{J}$ and $n_{J}$ are the spherical Bessel functions and $\eta_{J}(k)$ is the phase shift of the $J$-partial wave. 
The $J$-partial wave scattering parameter ($a_J$) is defined as:
\begin{equation}\label{eq:aJ}
    a_J = - \lim_{k\to 0} \frac{\tan{\eta_{J}(k)}}{k^{2J+1}}.
\end{equation}

Szmytkowski suggested another form of $a_J$ with an addition $(2J-1)!!(2J+1)!!$ prefactor to the above expression~\cite{Szmytkowski1995},
but the present work is based on Eq.~(\ref{eq:aJ}) for consistency with previous studies~\cite{Jamieson1998}. 
Eq.~(\ref{eq:aJ}) is only valid if the asymptotic potential satisfies the condition:
\begin{equation}
    \lim_{R\to\infty} R^{2J+3} V(R) = 0.
    \label{R-restrict}
\end{equation}
The dominating potential for the nonrelativisitic contribution is the dispersion energy with $R^{-6}$ term. 
The $s$-wave $J=0$ scattering parameters are well defined under this restriction.

The zero-energy radial wavefunction is evaluated by integrating Eq.~(\ref{eq:SE_transf}) outward at energy of $10^{-20}$ a.u. up to $R=500$ a.u. with 0.001 a.u. step size. 
The phase shift $\eta_{J}$ is derived using: 
\begin{equation}\label{CRH:eq:phase_shift_tan}
    \tan{\eta_{J}} = \frac{ Kj_{J}(kR_\text{a}) - j_{J}(kR_\text{b}) }{ Kn_{J}(kR_\text{a}) - n_{J}(kR_\text{b}) }; ~ K=\frac{ R_\text{a}\chi_{J}(R_\text{b}) }{ R_\text{b}\chi_{J}(R_\text{a})  },
\end{equation}
where the two outermost grid points $R_a = 499.999$ and $R_b = 500$ a.u.

The $s$-wave scattering length has been evaluated using this method in our previous work~\cite{Lai2022}, where $a_s$ was computed from the potential energy curves. There, larger differences were found for various levels of approximation (Born-Oppenheimer, adiabatic, non-adiabatic, relativistic), similar as in the work of Jamieson and Dalgarno~\cite{Jamieson1998}. 

The relativistic and QED contribution to $a_s$ could be evaluated using the perturbative approach presented in Ref.~\cite{Jamieson1998}. The correction term $\delta a_s$ in atomic units is given by:
\begin{equation}
    \delta a_s = \frac{2\mu_a}{k}\int_0^{R_b} \chi^2(R;k) \mathcal{E}^{(n,0)}(R) dR.
\end{equation}
The potential energy curves $\mathcal{V}(R)$ and $\mathcal{E}^{(4,0)}(R)$ are scaled by the same factor $f$ to evaluate $D^0_{14}(f)$ and $a_s(f)$. A simple scaling of the $W_{\|}(R)$ and the $ W_{\perp}(R)$ function will lead to nonphysical behavior of the vibrational and rotational reduced mass at $R\rightarrow 0$. 
The scaling of the $W_{\|}(R)$ and the $ W_{\perp}(R)$ functions has negligible effect on $a_s$ in the region of interest compared to the other contributions. $W_{\|}(R)$ and $ W_{\perp}(R)$ functions used in Eq.~(\ref{eq:SE_transf}) are kept unchanged in the current study.

\begin{figure}
\begin{center}
\includegraphics[width=\linewidth]{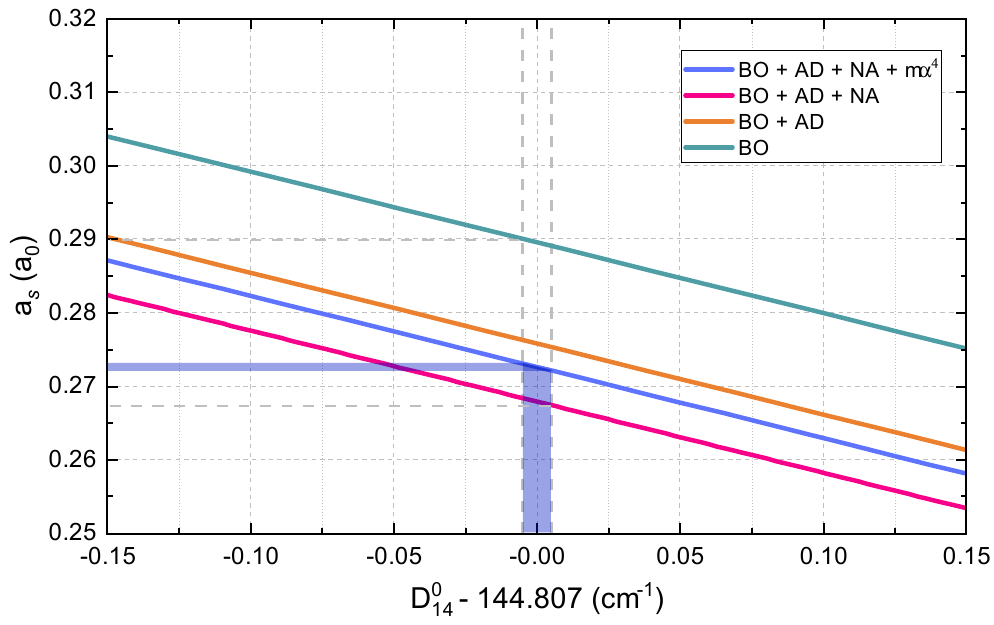}
\caption{\label{slen}
Relationship of the $s$-wave scattering length ($a_s$) with the binding energy of \Hm\ at $v=14, J=0$ state in the ground electronic state evaluated at different level of approximation. The yellow shaded region represents the extracted scattering length from the binding energy determined in this work. The grey dashed line represents the largest uncertainty within the given available approximation. 
}
\end{center}
\end{figure}

Figure \ref{slen} shows the correlation between $a_J$ and $D^J_{14}$ by varying $f$ at different level of approximations for $s$-wave scattering parameters. 
The slopes in Fig.~\ref{slen} are centered around the experimental value for the dissociation energy covering different ranges of $f$ factors. For example, the plotted range of $f$ for the $s$-wave scattering length at BO and AD level are about $[1.00098, 1.00109]$ and $[1.00045, 1.00056]$, respectively.

The numerical relation between $a_s$ and $D^0_{14}$ is linearly fitted within a narrow window of $D^0_{14}$. The slope of the fitting is fairly identical for all approximations as the $a_s$ and the $D^0_{14}$ heavily depend on the BO potential energy. The linear function is used to derive $a_s$ from the $D^0_{14}$ experimental value obtained in this work. These  values for $a_s$, as directly derived from experiment, at different level of approximations are listed in Table~\ref{tab:slen}.
\begin{table}
\caption{\label{tab:slen}
Values for the $s$-wave scattering length, obtained for the present direct method $a_s$(exp) and from the previous theoretical approach~\cite{Lai2022}, and  at different levels of approximation: BO - Born-Oppenheimer; AD - Adiabatic; NA - Nonadiabatic, and the $m\alpha^{n}$ contribution are listed. All values in units of $a_0$.
}
\renewcommand{\arraystretch}{1.3}
\begin{tabular}{cccc}
      & $a_s$ (exp) &  $a_s$ (theory~\cite{Lai2022}) \\
\hline
BO  &  0.2894 (5)  &  0.5698            \\
AD  &  0.2760 (5)  &  0.4159            \\
NA  &  0.2675 (5)   &  0.2572          \\
$m\alpha^{4}$ &  0.2724 (5)  & 0.2728      \\
\hline
\end{tabular}
\end{table}

The $\mathcal{E}^{(4,0)}(R)$, $\mathcal{E}^{(5,0)}(R)$ and $\mathcal{E}^{(6,0)}(R)$ potential energy curves are fitted with a series of $1/R^n$, where the leading order $n$ is 4, 3 and 2 respectively~\cite{Puchalski2016,Puchalski2017,Silkowski2023}. 
In the case of $s$-wave scattering, the $\mathcal{E}^{(5,0)}(R)$ and $\mathcal{E}^{(6,0)}(R)$ terms will violate the restriction imposed by Eq.~(\ref{R-restrict}).
Nonstandard treatment is necessary to include the contribution of these terms, yet they are of small a small amount and such treatment is beyond the scope of this work.

The uncertainties presented in Table~\ref{tab:slen} only account for the uncertainty in the experimental value of the dissociation energy and do not take into account uncertainties from the potential energy curves and their computation.
The $a_s$ extracted using the BO potential energy deviates about 6\% from the that with the leading order relativistic $m\alpha^4$ term. 
The semi-empirical method of Ref.~\cite{Augusticova2021} shows to be rather insensitive to small details of the potential energy curve and therewith forms a robust method to extracting a value for the scattering length.

\section{$p$-wave scattering Length}

\begin{figure}
\begin{center}
\includegraphics[width=\linewidth]{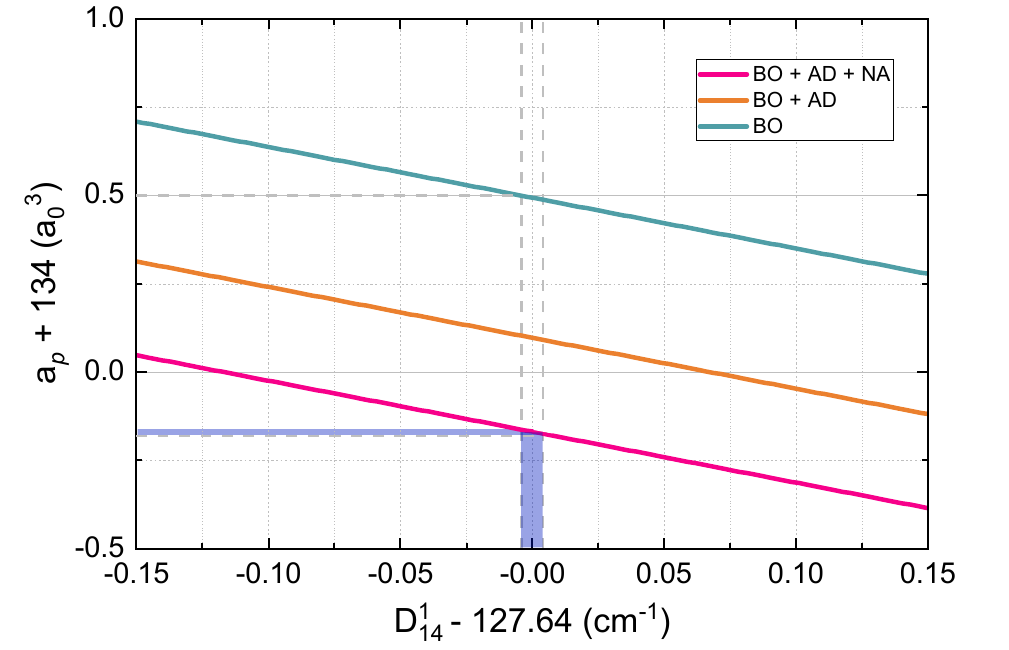}
\caption{\label{pvol}
Relationship of the $p$-wave scattering length ($a_p$) with the binding energy of \Hm\ at $v=14, J=1$ state in the ground electronic state evaluated at different level of approximation. The yellow shaded region represents the extracted scattering length from the binding energy determined in this work. The grey dashed line represents the largest uncertainty within the given available approximation.
}
\end{center}
\end{figure}

We have evaluated the $p$-wave scattering volume following similar procedures and applying the same boundary conditions and step size as in the evaluation of the $s$-wave scattering length. However, the evaluation of $p$-wave scattering volume requires special care of numerical accuracy of the algorithm. In this work, the $p$-wave scattering volume is evaluated using 25 decimal places of precision in the computation. 
Table~\ref{tab:pvol} lists the calculated $p$-wave scattering volume. 
The calculated scattering volume at the BO and adiabatic level of approximation agree well with the value ($a_p = -136.8$ $a_0^3$) obtained by Jamieson and Dalgarno~\cite{Jamieson2012}.

The relativistic and QED effects to the scattering volume cannot be evaluated using the perturbative approach as all terms decay slower than $R^{-5}$ in Eq.~(\ref{R-restrict}). 
Table~\ref{tab:pvol} lists the relativistic and QED effects to $p$-wave scattering volume using the perturbative approach. Since the asymptotic zero-energy wavefunction is proportional to $R^{2}$ for $J=1$, the calculated corrections by $m\alpha^{4-6}$ terms contribute significantly but its contribution is unphysical within the current model. 
Therefore, the relation of $D^1_{14}(f)$ and $a_p(f)$ is evaluated within the nonrelativsitic approximation only.

Figure~\ref{pvol} shows the relation of $D^1_{14}(f)$ and $a_p(f)$ at BO, adiabatic and nonadiabatic approximations. As listed in Table~\ref{tab:pvol}, the relative variation of unscaled scattering volume among different approximations is much smaller than the $s$-wave scattering length. However, a similar range of scaling factors $f$ are applied to shift the $D^1_{14}(f)$ close to the experimental value. 
The range of $f$ for the $p$-wave scattering length plotted in Fig.~\ref{pvol} at BO, AD and NA level are about $[1.00098, 1.00109]$, $[1.00045, 1.00056]$ and $[0.9999, 1.0001]$ , respectively. 
The $a_p$ at different stages are extracted from the linear fit within the presented window of $D^1_{14}$.

\begin{table}
\caption{\label{tab:pvol}
Values for the and $p$-wave scattering parameters (volumes), with $a_p$ (exp) via the direct method presented here and  $a_p$ (theory) determined via calculation via the NAPT procedure~\cite{Lai2022}, at different levels of approximation: BO - Born-Oppenheimer; AD - Adiabatic; NA - Nonadiabatic contributions are listed. All values in units of $a_0^3$.
}
\renewcommand{\arraystretch}{1.3}
\begin{tabular}{cccc}
      &     & $a_p$ (exp) & $a_p$ (theory) \\
\hline
BO  &     & -133.4991 (6)     & -129.3810       \\
AD  &     & -133.8947 (6)     & -131.8728       \\
NA  &     & -134.0000 (6)     & -134.1640       \\
\hline
\end{tabular}
\end{table}

\section{The last bound level: X, $v=14$, $J=4$}

In the literature there have been ample discussions as to whether the $X(14,4)$ rovibrational level of the electronic ground state of molecular hydrogen is a bound or a quasi-bound state~\cite{Komasa2011,Selg2011,Selg2012,Roueff2022}. 
In early vacuum ultraviolet emission spectroscopic studies decay to this state was observed as narrow line features~\cite{Dabrowski1984,Roncin1994}, but this is not decisive for settling the argument. 
In the present experimental study, via the recording of Fig.~\ref{F0X14Q4} and its subsequent analysis, a binding energy of 0.023 (4) \wn\ was determined, corresponding to 690 (120) MHz.
This value is in agreement with the result of Komasa et al.~\cite{Komasa2011} at 0.026 \wn.  We also reproduced a value in the computations in the framework of section~\ref{swave} within the error margins of the experiment.

\begin{figure}
\begin{center}
\includegraphics[width=\linewidth]{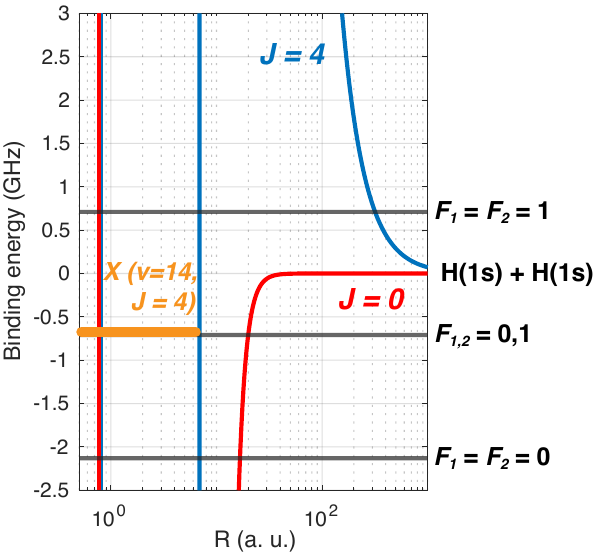}
\caption{\label{fig_limit}
Potential energy curves for $J=0$ (red) and $J=4$ (blue) of the electronic ground state X$^1\Sigma_g^+$. Taking the hyperfine splitting of the atomic fragments into account with $F_i=0,1$, three distinct H(1s) + H(1s) dissociation limits emerge, indicated by gray lines. 
The experimentally determined binding energy of X$(14,4)$, at 0.023 (4) \wn\ is indicated in orange, the bar representing the uncertainty.
}
\end{center}
\end{figure}

Harriman et al.~\cite{Harriman1967} have given a detailed analysis of the level structure near the dissociation threshold, that is very much different from the situation of the H$_2$ molecule at short internuclear separation. 
At short distance the X($v=14,N=J=F=4$) state has para-hydrogen character, where the nuclear spin $I=0$ does not result in any hyperfine splitting.
However, going toward larger $R$, the electron and nuclear spins of both hydrogen atoms become decoupled.
This decoupling can be understood as a breakdown of the para-ortho dichotomy, or the breaking of $g/u$ symmetry.
In the regime of intermediate internuclear separations, $R = 7-10$ $a_0$, the  X($N=J=F=4$) state will mix with the $N=4,F=4$ levels of the b$^3\Sigma_u^+$ state with $J=3,4,5$.
While the b$^3\Sigma_u^+$ potential has a minimum of only 4.3 \wn\ at  $R = 7.85$ $a_0$~\cite{Kolos1965}, insufficient to support bound states, this does not prohibit the aforementioned singlet-triplet mixing.
Finally, at very large $R$ with full separation of the atoms, effectively a hyperfine triplet results,  three dissociation limits with $F_1=F_2=1$, $F_{1,2}=0,1$, and $F_1=F_2=0$ as depicted in Fig.~\ref{fig_limit}.

Selg also discussed how the hyperfine structure, through the splitting of the dissociation limit into three separate limits, plays a role in the binding of the X(14,4) level~\cite{Selg2011,Selg2012}.
The hyperfine splitting in the H-atom amounts to 0.0473796 \wn~\cite{Karshenboim2005}.
With $F_i=0,1$ hyperfine states in the hydrogen atom, there are 16 possible $m_F$ hyperfine substates for the H$_2$ molecule at large internuclear separation: one for $F_1=F_2=0$, 6 for $F_{1,2}=0,1$ and 9 for $F_1=F_2=1$. 
The potential energy curves for \X~for $J=0$ and $J=4$, converging to the hyperfine-free dissociation limit, are shown in Fig.~\ref{fig_limit}. 
The three distinct dissociation thresholds are indicated by gray lines.
The degeneracies of the hyperfine combinations are accounted for in shifting the the center-of-gravity for H(1s)+H(1s), lying at $36\,118.069\,605$ (31) \wn\ above the ground rovibrational level of H$_2$~~\cite{Beyer2019}, to the zero level in the figure.
The energy values of the three hyperfine limits with respect to the hyperfineless limit are located at +0.0236898 \wn, at -0.0236898 \wn, and at -0.0710694 \wn. 

From the perspective of the three hyperfine dissociation limits the actual X(14,4) level lies above the lowest limit ($F_1=F_2=0$), and below the upper limit ($F_1=F_2=1$), while it coincides, within uncertainty limits with the internediate limit ($F_{1,2}=0,1$).
This suggests, that the level is actually turned into a Feshbach resonance. 

Selg~\cite{Selg2011} attempted the calculation of the correct level position including effects of nuclear spins, and indeed found that the bound state was turned into a resonance or quasi-bound state, with a lifetime (5 minutes) too long to lead to any noticible difference in the spectra presented here. While Selg did not account for the full hyperfine splitting of all three involved disssociation limits, this correction would probably not lead to any noticeable difference at the level of the current resolution.

\section{Conclusion}

The two-photon UV-laser photolysis of hydrogen sulfide (H$_2$S) has become a benchmark platform for the investigation of highly excited rovibrational quantum states in the X$^1\Sigma_g^+$ electronic ground state of the hydrogen molecule in the vicinity of the dissociation threshold.
Previously, this minor dissociation channel has allowed for the spectroscopic analysis of high-lying vibrational levels ($v=7-13$) for wide ranges of rotational levels. In these studies the experimental determinations of level energies were found to be in excellent agreement with advanced theoretical computations including subtle effects of relativistic quantum electrodynamics. 
Thereupon the energy range above the H+H+S($^1$D) dissociation threshold was explored and a number of scattering resonances were observed as quasi-bound molecular states via the same three-step laser excitation schemes that was explored for the bound resonances.

In the present study the focus is on the five rotational states in the $v=14$ vibrational manifold of \Hm(X). In particular the observation and precise measurement of the $J=0$ and $J=4$ levels are of great interest.
The binding energy of the $J=0$ level, at 144.807 (5) \wn, can be converted via a direct spectroscopic approach~\cite{Augusticova2021} to yield an $s$-wave scattering length for H+H collisions at $a_s = 0.2724\,(5)$ a$_0$. 
Finally, the binding energy of the last bound level in H$_2$, the ($v=14,J=4$) level, was experimentally determined to lie 0.023 (4) \wn\ below the hyperfineless dissociation limit. 
The discussion in the literature whether this $J=4$ level is bound or unbound~\cite{Komasa2011,Selg2011,Selg2012,Roueff2022} is elucidated. The level is bound with respect to the $F_1=F_2=1$ limit including atomic hyperfine structure, well above the $F_1=F_2=0$ hyperfine limit, while it coincides, within experimental and theoretical uncertainty limits, with the $F_{1,2}=0,1$ hyperfine limit.

Finally we remark that in the previous experimental~\cite{Cheng2018,Holsch2019,Hussels2022,Holsch2023} and theoretical~\cite{Puchalski2018,Puchalski2019} studies determining accurate  dissociation limits of the hydrogen molecule and its isotopologues the convention was followed to report values with respect to the hyperfineless H(1s)+H(1s) limit. 
When analyzing the dynamics of bound levels close to the dissociation limit the effect of hyperfine structure of the atomic fragments must be taken into account.


%

\end{document}